\documentclass[a4paper, amsfonts, amssymb, amsmath, showkeys, nofootinbib]{revtex4-2}
\usepackage[svgnames]{xcolor}
\usepackage[left=23mm,right=13mm,top=35mm,columnsep=15pt]{geometry} 
\usepackage{placeins}
\usepackage{enumitem}
\usepackage{hhline}
\newcommand{\specialcell}[2][c]{%
	\begin{tabular}[#1]{@{}c@{}}#2\end{tabular}}
\usepackage{algorithm}
\usepackage[rightComments=true, noEnd=false, italicComments=false, commentColor=Green, endLComment= ]{algpseudocodex}
\usepackage[pdftitle={Article}, pdfauthor={Author}]{hyperref} 
\begin{document}
\raggedbottom
\title{NonSysId: A nonlinear system identification package with improved model term selection for NARMAX models}

\author{Rajintha Gunawardena\textsuperscript{1}}
\author{Zi-Qiang Lang\textsuperscript{2}}
\author{Fei He\textsuperscript{1}}
    \email[Correspondence email address: ]{fei.he@coventry.ac.uk}
\affiliation{
\begin{enumerate}
    \item Centre for Computational Science and Mathematical Modelling, Coventry University, Coventry CV1 5FB, UK
    \item School of Electrical and Electronic Engineering, The University of Sheffield, Western Bank, Sheffield S10 2TN, UK
\end{enumerate}
}
\begin{abstract}
\section*{Summary}
System identification involves constructing mathematical models of dynamic systems using input-output data, enabling analysis and prediction of system behaviour in both time and frequency domains. This approach can model the entire system or capture specific dynamics within it. For meaningful analysis, it is essential for the model to accurately reflect the underlying system's behaviour. This paper introduces NonSysId, an open-sourced MATLAB software package designed for nonlinear system identification, specifically focusing on NARMAX models. The software incorporates an advanced term selection methodology that prioritises on simulation (free-run) accuracy while preserving model parsimony. A key feature is the integration of iterative Orthogonal Forward Regression (iOFR) with Predicted Residual Sum of Squares (PRESS) statistic-based term selection, facilitating robust model generalisation without the need for a separate validation dataset. Furthermore, techniques for reducing computational overheads are implemented. These features make NonSysId particularly suitable for real-time applications such as structural health monitoring, fault diagnosis, and biomedical signal processing, where it is a challenge to capture the signals under consistent conditions, resulting in limited or no validation data. 

\noindent \textbf{URL:} \href{URL}{https://github.com/raj-gun/NonSysID}
\end{abstract}
\maketitle
\section{Statement of need}
System identification is a field at the intersection of control theory, dynamic systems theory and machine learning that seeks to derive mathematical models of dynamic linear or nonlinear systems based on experimental input-output data. 
Generally, system identification has two primary objectives \cite{ljung1998system,billings2013a}, (i) to accurately map the system's inputs and outputs, allowing for the prediction of new, unseen data, and (ii) to capture the underlying dynamics of the system within the model.

The dynamic models generated through system identification can be either discrete or continuous time models \cite{UNBEHAUEN1997}. This paper centers on widely-used discrete-time models, specifically nonlinear auto-regressive models with exogenous inputs (NARX), where the ARX model is a linear variant of the NARX framework. These input-output time-series models predict future outputs of a system based on its historical input and output instances. NARX models have been applied extensively to model and analyse complex systems in fields such as control, fault diagnosis, structural health monitoring and the modelling and analysis of physiological and biological systems \cite{Chiras2002,WANG2024,ZAINOL2022,RITZBERGER2017,Gao2023,HE2016,HE2021}. Moreover, it has been demonstrated that the NARX model has equivalence to a recurrent neural network (RNN) \cite{Sum1999}. Extending the NARX model to incorporate a noise model, we obtain the nonlinear auto-regressive moving average with exogenous inputs (NARMAX) model.  

Recently two open-sourced packages have been introduced, `SysIdentPy' \cite{Lacerda2020} for Python and the `narmax' package \cite{AYALA2020} for R. Both packages are well-developed and comprehensive. However, they are based on the original forward regression orthogonal least squares (OFR) algorithm, which has been noted to have several limitations, as discussed in \cite{Piroddi2003,Mao1997}. These concerns primarily involve over-fitting and inaccurate long-horizon predictions, particularly when the input fails to sufficiently excite the actual system. Additionally, in some applications, acquiring extra data for cross-validation may be infeasible. As a result, developing parsimonious models that can generalise well to unseen data becomes crucial in such cases. This paper introduces the `NonSysId' package, which incorporates an enhanced model selection process to address these challenges.

In the context of (N)ARX models, system identification is employed to determine a specific functional relationship that maps past input instances (input-lagged terms),
\begin{equation}\label{eq:Ut_sysid}
    U = \Big\{ u(t-1)\ ,\ u(t-2)\ ,\ \cdots,\ u(t-n_b) \Big\},
\end{equation}
and past output instances (output-lagged terms),
\begin{equation}\label{eq:Yt_sysid}
    Y = \Big\{ y(t-1)\ ,\ y(t-2)\ ,\ \cdots,\ y(t-n_a) \Big\}, 
\end{equation}
to the present output instance in time $y(t)$. $t$ here refers to a time index (i.e. $t$\textsuperscript{th} sample). $n_a$ and $n_b$ are the maximum number of past output and input time instances considered and are related to the Lyapunov exponents of the actual system that is being modelled \cite{mendes1998a}. The functional mapping is described by the following equation:
\begin{equation}\label{eq:sys_id_func}
y(t) = f^{P}\bigl( Y, U \bigr) + \xi(t),
\end{equation}
where $y(t)$ and $u(t)$ refer to the output and input respectively, while $\xi(t)$ represents the error between the predicted output $f^{P}\bigl( Y, U \bigr)$ and the actual output $y(t)$ at time instance $t$. $\xi(t)$ will contain noise and unmodeled dynamics. $f^{P}( \ )$ is the functional mapping between the past inputs and outputs to the current output $y(t)$. This mapping can take the form of a polynomial, a neural network, or even a fuzzy logic-based model. Here, we focus on polynomial NARX models with a maximum polynomial degree $N_p \in \mathbb{Z}^{+}$. In this case, Eq. \eqref{eq:sys_id_func} can be expressed as
\begin{equation}\label{eq:sys_id_func_summation}
y(t) = \sum_{m=1}^{M} \theta_{m} \times \phi_{m}(t) + \xi(t),
\end{equation}
where $m = 1, \cdots, M$, $M$ being the total number of variables or model terms. $\theta_{m}$ are the model parameters or coefficients and $\phi_{m}(t)$ are the corresponding model terms or variables. $\phi_{m}(t)$ are $n$\textsuperscript{th}-order monomials of the polynomial NARX model $f^{P}( \ )$, where $n = 1, \cdots, N_p$ is the degree of the monomial. $\phi_{m}(t)$ is composed of past output and input time instances from $Y$ and $U$. An example of a polynomial NARX model can be
\begin{equation}\label{eq:narx_exmpl}
    y(t) = \theta_{1}y(t-1) + \theta_{2}u(t-2) + \theta_{3}y(t-2)^{2}u(t-1)^{3} + \xi(t).
\end{equation}
In this example, $\phi_{1}(t)=y(t-1)$ and $\phi_{2}(t)=u(t-2)$ have a degree of 1 and are the linear terms (1\textsuperscript{st} order monomials or linear monomials) of the model. $\phi_{3}(t) = y(t-2)^{2}u(t-1)^{3}$ is a nonlinear term with a degree of $5$ (5\textsuperscript{th} order monomial, more generally a nonlinear monomial). The NARX model given in Eq. \ref{eq:narx_exmpl} has a polynomial degree $N_p=5$ (highest degree of any monomial). Given that the total number of time samples available is $L$, where $t = 1, \cdots, L$, Eq. \ref{eq:sys_id_func_summation} can be represented in matrix form as
\begin{equation}\label{eq:sys_id_func_mat}
\mathbf{Y} = \mathbf{\Phi} \mathbf{\Theta} + \mathbf{\Xi},
\end{equation}
where $\mathbf{Y} = \left[ y(1), \cdots, y(L) \right]^T$ is the vector containing the output samples $y(t)$. $\mathbf{\Phi} = \left[ \bar{\phi}_{1}, \cdots, \bar{\phi}_{M} \right]$, where $\bar{\phi}_{m} = \left[ \phi_{m}(1), \cdots, \phi_{m}(L) \right]^T$ is the vector containing all time samples of the model term $\phi_{m}(t)$. $\mathbf{\Theta} = \left[ \theta_{1}, \cdots, \theta_{M}  \right]^T$  is the parameter vector and $\mathbf{\Xi} = \left[ \xi(1), \cdots, \xi(L) \right]$ is the vector containing all the error terms $\xi(t)$ (i.e. model residuals). In the NARMAX model structure, a moving-average (MA) component is added to the NARX (Eq. \eqref{eq:sys_id_func_summation}) by incorporating linear and nonlinear lagged error terms (e.g., $\xi(t-2)$, $\xi(t-1)\xi(t-3)$). This noise model accounts for unmodeled dynamics and coloured noise, effectively isolating noise from the deterministic system and thereby reducing model bias \cite[Chapter~3]{billings2013a}.

The primary challenge in learning a polynomial NARX model is to identify the polynomial structure of the model, i.e. selecting which terms from a set of candidate model terms (monomials), denoted as $\mathcal{D}$, should be included in the model. For instance, a potential set of candidate terms could be
\begin{equation}\label{eq:exmpl_D}
    \mathcal{D} = \Big\{ 
              y(t-1), y(t-2), u(t-1), u(t-2), 
              y(t-1)u(t-2), y(t-2)u(t-1)^{3}, 
              y(t-2)^{2}u(t-1), y(t-2)^{2}u(t-1)^{3}
        \Big\} ,
\end{equation}
from which a NARX model structure, such as that in Eq. \ref{eq:narx_exmpl}, can be identified. Once the model structure is identified, the next step is to estimate the model parameters. However, determining the appropriate linear and nonlinear terms to include in the model structure is critical to achieving parsimonious models. This is particularly important in the nonlinear cases \cite[Chapter~1]{billings2013a}, as the inclusion of  unnecessary model terms, can result in a model that erroneously captures dynamics that do not belong to the underlying system \cite{AGUIRRE1995, mendes1998a}.

The Orthogonal Forward Regression (OFR) algorithm, also known as Forward Regression OLS (FROLS) \cite{chen1989b, billings1987a}, is based on the Orthogonal Least Squares (OLS). When combined with an appropriate term selection criterion \cite{korenberg1988a, WANG1996, hong2003}, it efficiently selects model terms (regressors) in a forward, sequential manner. In this approach, model terms are added one at a time based on a selection criterion, facilitating the development of a parsimonious model. The OFR/FROLS algorithm evaluates the impact of each term on the output independently of the influence of other terms, achieved through  orthogonalization procedures. This evaluation relies on the chosen term selection criterion, allowing for the sequential inclusion of appropriate terms in the final model using a forward selection approach. The most commonly used and widely accepted model term selection criterion used in the OFR algorithm is the error reduction ratio (ERR). During the forward selection procedures, the ERR selects the term that maximises explained variance, thereby  maximise the goodness of fit. Over the years, many variants of the OFR have been proposed. However, concerns persist regarding the original OFR algorithm (OFR-ERR), which relies on the ERR for term selection \cite{Piroddi2003,Mao1997}, for example,
\begin{enumerate}
    \item OFR-ERR may select redundant or incorrect model terms, especially in the presence of complex noise structures or certain input signals;
    \item The model structures produced from OFR can be sensitive to the first term selected in the forward selection of model terms;
    \item If the input does not persistently excite the system under consideration (i.e. it lacks the informativeness needed to effectively stimulate the system), the resulting model can be inappropriate. This can result in inaccuracies in long-horizon prediction and, in some cases, even unstable models during simulation (free-run or model-predicted output);
    \item The ERR focuses solely on explained variance when selecting terms, which can lead to overfitting. 
\end{enumerate}

Beyond obtaining parsimonious models, the model should generalise well to unseen data (validation) that is not used during the learning/training process (i.e. model identification). This is referred to as obtaining a bias-variance trade-off, which can be achieved through an appropriate cross-validation strategy \cite{Little2017,Stone1974}. However, in some applications, obtaining separate validation data is not feasible. This is particularly true in real-time system identification applications, such as structural health monitoring or fault diagnosis \cite{Gharehbaghi2022,Vamsikrishna2024}. Another example arises in neuroscience, where the dynamics between brain regions are highly time-varying and can change within milliseconds. As a result, obtaining electrophysiological data that precisely captures such behaviour is often challenging, if not impossible \cite{Kunjan2021,Seedat2024,Chen2016,Eichenbaum2021,Lehnertz2021}. These challenges are critical when applying system identification to specific domains. The following section outlines the features in the `NonSysId' package, designed to address these issues.   
\section{Features in NonSysId}
The `NonSysId' package introduced in this paper implements an OFR-based system identification methodology designed to addresses the key issues mentioned in the latter part of the previous section. This is achieved by integrating and extending several OFR variants already available in the literature \cite{guo2015a,WANG1996, hong2003}, along with a proposed simulation-based model selection procedure. 
A notable feature of `NonSysId' is the implementation of the iterative-OFR (iFRO) variant \cite{guo2015a} of the OFR algorithm. Additionally, the PRESS-statistic-based term selection \cite{WANG1996, hong2003} is integrated with the iOFR, complemented by simulation-based model selection. These enhancements enable robust term selection (compared to the ERR), built-in cross-validation, and the ability to produce models with long-horizon prediction capabilities and simulation stability \cite{AGUIRRE2010}. With these features, the `NonSysId' package makes system identification feasible for real-time applications, such as fault diagnosis in engineering or the analysis of electrophysiology activity in medical settings, where inputs may not be persistently exciting and separate datasets for validation may be unavailable. 
`NonSysId' is the only open-sourced package that directly address the limitations of the original OFR algorithm. For NARX models, where the candidate term set can be extensive and computationally demanding in the iFRO algorithm, `NonSysId' incorporates methods to reduce the candidate term set, significantly speeding up the forward selection process. Moreover, the package includes correlation-based residual analysis techniques for nonlinear model validation \cite{Billings1983}.
\subsection{Iterative OFR}\label{sec:iOFR}
To address the concerns associated with the original OFR, the iterative-OFR (iOFR) algorithm was introduced in \cite{guo2015a}. To the best of our knowledge, no open-source software implementing this variant currently exist. In the original OFR, the term selection is heavily influenced by the order of orthogonalization, which can often result in incorrect terms being selected in the early stages \cite{guo2015a, Mao1997}. Additionally, the order in which terms are selected in the OFR determines orthogonalization path, resulting in a tree structure of possible models \cite{guo2015a, Mao1997}. Finding a globally optimal solution would require an exhaustive search through all orthogonalization paths - an infeasible task given the factorial  growth in paths ($k!$ paths for $k$ terms). The iOFR algorithm addresses this limitation by iteratively exploring multiple orthogonalization paths and re-selecting terms to approximate a globally optimal model without exhaustive search \cite{guo2015a}. This approach enables the recovery of correct terms that might have been overlooked in earlier iterations. As a result, the iOFR generates several candidate models for consideration. 

The iOFR procedures \cite{guo2015a} can be summarised as follows. Given an output vector $\mathbf{Y}$, a set of candidate terms $\mathcal{D}$ and a set of pre-select terms $\mathcal{P} \subseteq \mathcal{D}$, where $\mathcal{P} = \{ \phi_1 , \dots, \phi_p  \}$, 1) pre-select each term given in $\mathcal{P}$ as the first model term; 2) use OFR to search through $p$ orthogonalization paths resulting in a set of $p$ candidate models $\mathcal{M} = \{ m_1, \dots, m_p \}$; 3) from $\mathcal{M}$, select the best model $\overline{m}$ based on the one-step-ahead prediction error; and 4) update the set of pre-select terms $\mathcal{P}$ with the terms in $\overline{m}$. The process is repeated iteratively with the updated $\mathcal{P}$ to search through different orthogonalization paths.    

As shown in \cite{guo2015a}, the iOFR can iteratively produce more globally optimal model structures. This is because optimal solutions are only found along orthogonalization paths that begin with a correct term \cite{guo2015a, Mao1997} (candidate terms essential for accurately reconstructing dynamics of the original system \cite{mendes1998a, AGUIRRE1994}). Although the best model $\overline{m}$ obtained in each iteration may be sub-optimal, it will include certain correct terms \cite{guo2015a}. Consequently, in subsequent iterations, $\mathcal{P}$ will contain fewer redundant model terms. This refinement ensures that, in the next iteration, relatively greater proportion of the orthogonalization paths explored by the OFR will start from better initial terms, leading to a more robust set of models $\mathcal{M}$ \cite{guo2015a}. For the first iOFR iteration, the pre-select terms, $\mathcal{P}$, can be set to $\mathcal{P} = \mathcal{D}$. Since it is sufficient to focus on orthogonalization paths that begin with correct terms \cite{guo2015a}, methods for selecting the initial set $\mathcal{P}$ will be discussed in later sections. This will make the iOFR converge faster towards an optimum while improving computational efficiency.

In the original iOFR algorithm \cite{guo2015a}, model selection was based on the one-step-ahead prediction. The implementation of iOFR in the `NonSysId' extends this by incorporating simulation-based model selection ($\text{iOFR}_{S}$) to ensure simulation stability and improve long-horizon prediction accuracy. The procedures for $\text{iOFR}_{S}$ are as follows:
\begin{enumerate}
    \item Pre-select each term given in $\mathcal{P}$ as the first model term and search through $p$ orthogonalization paths using OFR to produce a set of $p$ candidate models $\Tilde{\mathcal{M}} = \{ \Tilde{m}_1, \dots, \Tilde{m}_p \}$. 
    \item From $\Tilde{\mathcal{M}}$ determine the set of stable candidate models $\mathcal{M} = \{ m_1, \dots, m_{\overline{p}} \}$, $\overline{p} \leq p$.
    \item From $\mathcal{M}$, based on the simulation error choose the best model $\overline{m}$. 
    \item Use the terms in $\overline{m}$ to form the new set of pre-select terms $\mathcal{P}$.
    \item Repeat steps 1-4 and iteratively search through different orthogonalization paths.
\end{enumerate}
In step 2, each model $\Tilde{m}_i \in \Tilde{\mathcal{M}}$, $i=1,\dots,p$, is tested using two inputs: (i) a sequence of 0's, $u^{[0]}(t) = 0 \ \forall t$, and (ii) a sequence of 1's, $u^{[1]}(t) = 1 \ \forall t$. The corresponding simulated outputs, $\hat{y}^{[0]}(t)$ and $\hat{y}^{[1]}(t)$, must meet stability conditions for $\Tilde{m}_i$ to be included in $\mathcal{M}$. In this context, stability implies that the outputs remain bounded over time, i.e. stable around a mean without exhibiting exponential growth. Specifically, the responses ($j = 0 \ \text{or} \ 1$) should be around a mean, $\mathbb{E}[\hat{y}^{[j]}(t)] \in \mathbb{R}$, with a small variance, $\text{Var}(\hat{y}^{[j]}(t)) \leq \varepsilon$. Typically, $\varepsilon = 10^{-2}$. Specifically, for $u^{[0]}(t)$, $\mathbb{E}[\hat{y}^{[0]}(t)] = \beta$, where $\beta$ is the bias term (DC offset) in the model, with $\beta=0$ indicating the absence of a bias term. In step 3, the Bayesian Information Criterion (BIC) \cite{Schwarz1978, Stoica2004} is used to select the optimal model $\overline{m}$ from $\mathcal{M}$. The BIC is calculated based on the simulated error variance (i.e. mean squared simulated error \cite{Piroddi2003}—MSSE) between the actual output and the model's simulated output. The $\text{iOFR}_{S}$ can be represented in functional form as $( \mathcal{M}, \overline{m} ) = \text{iOFR}_{S}( \mathcal{D}, \mathcal{P}, \mathbf{Y})$.
\subsection{PRESS-statistic-based term selection}\label{sec:PRESS}
The model must generalise effectively to unseen data during training, striking a balancing bias and variance. This can be achieved using robust cross-validation strategies. Ideally, an algorithm should optimise model generalisation without relying on a separate validation dataset. A PRESS-statistic-based \cite{Allen1974} term selection criterion with leave-one-out cross-validation was introduced into the OFR framework in \cite{WANG1996, hong2003}. Leveraging the OLS method in OFR, the computation of the leave-one-out cross-validation errors is highly efficient \cite{WANG1996, hong2003}. Integrating the PRESS-statistic-based criterion into the OFR algorithm enables the selection of regressors (model terms) that incrementally minimise the one-step ahead leave-one-out cross-validation error in a forward selection manner, effectively reducing overfitting to noise. This approach fully automates the model evaluation process, eliminating the need for additional validation data. Consequently, applying the PRESS-statistic-based term selection criterion within the $\text{iOFR}_{S}$ algorithm enhances the selection of more robust terms and improves the generalisation capabilities of the resulting models.
\subsection{$\text{iOFR}_{S}$ with reduced computational time}
This section outlines the comprehensive procedures implemented in the `NonSysId' package for identifying (N)ARX models from system input-output data using $\text{iOFR}_{S}$ algorithm combined with PRESS-statistic-based term selection. Additionally, techniques for reducing computational time in $\text{iOFR}_{S}$ are discussed. These techniques focus on efficiently reducing the number of candidate terms, pre-select terms, or both.       

Let $\mathcal{D'}$ denote the set of candidate linear terms comprising past inputs $U$ and outputs $Y$ (as defined in Eq. \eqref{eq:Ut_sysid} and \eqref{eq:Yt_sysid}, respectively), such that $\mathcal{D'} = Y \cup U$. This set is used to identify or learn an ARX model. Similarly, let $\mathcal{D}''$ represent the set of candidate terms that includes both linear and nonlinear terms, enabling the identification of a NARX model. Typically, $\mathcal{D}''$ is constructed by expanding $\mathcal{D'}$ to include additional nonlinear terms (nonlinear monomials) generated through combinations of terms in $\mathcal{D}'$, such that $\mathcal{D}'' \supset \mathcal{D}'$. However, as the number of past inputs and outputs increases (i.e. number of terms in set $\mathcal{D'}$, $|\mathcal{D'}|$) and higher degrees of nonlinearity are considered, the number of candidate nonlinear terms, $|\mathcal{D}'' - \mathcal{D'}|$, can increase exponentially \cite{billings2013a}. This rapid growth significantly raises the computational time needed to build a NARX model using the iOFR or $\text{iOFR}_{S}$ algorithms. Therefore, reducing the candidate set $\mathcal{D}''$, $\mathcal{D}''_{R}$ ($\mathcal{D}''_{R} \subset \mathcal{D}''$), can significantly decrease the search space for model terms, thereby offering a computational advantage. Additionally, in the initial iteration of the iOFR algorithm, minimizing the presence of redundant terms in the pre-select set $\mathcal{P}$ (i.e. the initial set $\mathcal{P}$) can expedite convergence toward an optimum model \cite{guo2015a}, which also applies to $\text{iOFR}_{S}$. This section will explore methods for obtaining a reduced $\mathcal{D}''$, i.e. $\mathcal{D}''_{R}$, and a more effective initial set $\mathcal{P}$. The techniques presented aim to streamline the search space and reduce the computational demands of $\text{iOFR}_S$.  

A technique for obtaining a reduced set of candidate model terms, $\mathcal{D}''_{R}$, was proposed in \cite{Wei2004}. This approach is based on the idea that if a lagged term significantly influences the output of a nonlinear system, it will also be significant in a linearised representation of the system. Accordingly, a linear ARX model is identified first, serving as a linearised model of the actual nonlinear system. The terms from this ARX model are then used to construct $\mathcal{D}''_{R}$. This method has been incorporated into the `NonSysId' package. Regarding the set $\mathcal{P}$, an initialisation method for the iOFR algorithm was proposed in \cite{guo2015a}. In this method, an overfitting NARX model, $\overline{m_{0}''}$, is first identified using the OFR. The terms from $\overline{m_{0}''}$ are then used to construct the initial set $\mathcal{P}$ for the first iteration of the iOFR. While $\overline{m_{0}''}$ may be sub-optimal, it is likely to include some correct terms. Consequently, using the terms of $\overline{m_{0}''}$ to form the initial set $\mathcal{P}$ ensures fewer redundant terms compared to directly setting $\mathcal{P} \subseteq \mathcal{D}$ \cite{guo2015a}. 

The `NonSysId' package incorporates both aforementioned methods to reduce the computational time of $\text{iOFR}_{S}$. Additionally, the `NonSysId' package implements two new methods, proposed in this paper, to further enhance computational efficiency (referred to as reducing computational time, RCT, methods). Algorithm \ref{alg:NonSysId} outlines the procedures of the complete system identification methodology, integrating $\text{iOFR}_{S}$ with these four RCT methods. A brief overview of each RCT method is provided below.
\begin{description}
    \item[RCT Method 1] This method, as proposed in \cite{Wei2004}, seeks to obtain a reduced set of candidate model terms, enabling $\text{iOFR}_{S}$ to operate within a narrower search space defined by $\mathcal{D}''_{R} \subset \mathcal{D}''$.
    \item[RCT Method 2] This method, as proposed in \cite{guo2015a}, identifies an appropriate initial set of pre-select terms, $\mathcal{P}$, for the first iteration of the iOFR algorithm. By ensuring that $\mathcal{P} \subset \mathcal{D}''$ (contains fewer redundant terms compared to $\mathcal{P} \subseteq \mathcal{D}$), the first iteration of iOFR/$\text{iOFR}_{S}$ involves fewer orthogonalization paths originating from redundant terms. This accelerates convergence towards an optimal model \cite{guo2015a}.
    \item[RCT Method 3] This method combines RCT methods 1 and 2, such that $\mathcal{P} \subset \mathcal{D}''_{R}$ and $\text{iOFR}_{S}$ searches through an appropriately reduced space defined by $\mathcal{D}''_{R}$. As a result, this approach enables faster convergence of $\text{iOFR}_{S}$ to an optimal model compared to any other RCT method.
    \item[RCT Method 4] This method combines RCT methods 1 and 2, such that $\mathcal{P} \subset \mathcal{D}''_{R}$. However, $\text{iOFR}_{S}$ searches through the full space $\mathcal{D}''$ instead of $\mathcal{D}''_{R}$. Therefore, this technique converges the $\text{iOFR}_{S}$ faster to an optimal model compared to RCT method 2.    
\end{description}
\begin{algorithm}[H]
\caption{Identify (N)ARX model using $\text{iOFR}_{S}$}\label{alg:NonSysId}
\begin{algorithmic}[1]
\State \textbf{Input:} $U$, $Y$, $\mathbf{Y}$ 
\LComment{\textbf{Identify ARX model}}
\State Compose the set of candidate linear terms $\mathcal{D'} = Y \cup U$
\State $\mathcal{P} = \mathcal{D'}$
\Comment{Set of pre-selecting terms}
\State $( \mathcal{M}', \overline{m}' ) = \text{iOFR}_{S}( \mathcal{D}', \mathcal{P}, \mathbf{Y})$
\Comment {Apply $\text{iOFR}_{S}$ to obtain the best ARX model $\overline{m'}$ out of all the competing models $\mathcal{M}'$}
\LComment{\textbf{Identify NARX model}}
\If{a NARX model is to be identified}
    \State $\mathcal{D}'' = \mathcal{D}' \cup f(\mathcal{D}')$
    \Comment{$\mathcal{D}''$ is composed by expanding $\mathcal{D'}$ to include additional nonlinear terms. $\mathcal{D}'' \supset \mathcal{D}'$}
    \State $\mathcal{D}'_{m}$ = terms of $\overline{m'}$
    \Comment{Set containing the terms of the ARX model $\overline{m'}$. $\mathcal{D}'_{m} \subset \mathcal{D}'$}
    \State $\mathcal{D}''_{R} = \mathcal{D}'_{m} \cup f( \mathcal{D}'_{m} )$
    \Comment{Use $\mathcal{D}'_{m}$ to form a reduced version of $\mathcal{D}''$, $\mathcal{D}''_{R}$. $\mathcal{D}''_{R} \subset \mathcal{D}''$}
    \If{RCT is not required}
        \LComment{If RCT is not required, all possible orthogonalisation paths are searched}
        \State $\mathcal{P} = \mathcal{D}''$
        \Comment{Set of pre-selecting terms}
        \State $( \mathcal{M}'', \overline{m''} ) = \text{iOFR}_{S}( \mathcal{D}'', \mathcal{P}, \mathbf{Y})$
        \Comment {Apply $\text{iOFR}_{S}$ to obtain the best NARX model $\overline{m''}$ out of $\mathcal{M}''$}
    \ElsIf{RCT method 1}
        \LComment{\textbf{RCT Method 1}}
        \State $\mathcal{P} = \mathcal{D}''_{R}$
        \State$( \mathcal{M}'', \overline{m''} ) = \text{iOFR}_{S}( \mathcal{D}''_{R}, \mathcal{P}, \mathbf{Y})$
    \ElsIf{RCT method 2}
        \LComment{\textbf{RCT Method 2}}
        \State Apply the OFR algorithm on $\mathcal{D''}$ to obtain an overfitting NARX model $\overline{m_{0}''}$.
        \State $\mathcal{P}$ = terms of $\overline{m_{0}''}$
        \Comment{Set of pre-selecting terms}
        \State$( \mathcal{M}'', \overline{m''} ) = \text{iOFR}_{S}( \mathcal{D}'', \mathcal{P}, \mathbf{Y})$
    \ElsIf{RCT method 3}
        \LComment{\textbf{RCT Method 3}}
        \State Apply the OFR algorithm on $\mathcal{D''}_{R}$ to obtain an overfitting NARX model $\overline{m_{0}''}$.
        \State $\mathcal{P}$ = terms of $\overline{m_{0}''}$
        \State$( \mathcal{M}'', \overline{m''} ) = \text{iOFR}_{S}( \mathcal{D}''_{R}, \mathcal{P}, \mathbf{Y})$
    \ElsIf{RCT method 4}
        \LComment{\textbf{RCT Method 4}}
        \State Apply the OFR algorithm on $\mathcal{D''}_{R}$ to obtain an overfitting NARX model $\overline{m_{0}''}$.
        \State $\mathcal{P}$ = terms of $\overline{m_{0}''}$
        \State$( \mathcal{M}'', \overline{m''} ) = \text{iOFR}_{S}( \mathcal{D}'', \mathcal{P}, \mathbf{Y})$
    \EndIf
    \If{ BIC of $\overline{m'}$ $\leq$ BIC of $\overline{m''}$}
        \LComment{Compare ARX model $\overline{m'}$ and NARX model $\overline{m''}$ using the BIC based on \textit{simulation error variance}. This is to determine if the ARX model is enough to explain the input-output mapping}
        \State \textbf{Output:} $\mathcal{M}'$, $\overline{m}'$ 
    \Else
        \State \textbf{Output:} $\mathcal{M}'$, $\overline{m}'$, $\mathcal{M}''$, $\overline{m}''$ 
    \EndIf
\Else
    \State \textbf{Output:} $\mathcal{M}'$, $\overline{m}'$
\EndIf
\end{algorithmic}
\end{algorithm}

The RCT methods aim to accelerate the convergence of $\text{iOFR}_{S}$ and reduce the time required to obtain a model. Using $\mathcal{D}''_{R}$ reduces the computational time for the OFR algorithm within $\text{iOFR}_{S}$, by shortening the time needed to follow a given orthogonalization path. Additionally, fewer redundant terms in $\mathcal{P}$ lead to faster convergence of $\text{iOFR}_{S}$ and contribute to reducing time by minimizing the number of orthogonalization paths \cite{guo2015a}. Therefore, the most effective RCT method is 3, followed by methods 1,4 and 2. However, when reducing the search space (determining $\mathcal{D}''_{R}$), RCT methods 1 and 3 may miss some correct terms, potentially resulting in convergence to a sub-optimal model. This outcome depends on the level of white and coloured noise in the input-output data, as well as the complexity of the original system. It should be noted that RCT methods introduce additional procedures. Therefore, if $\mathcal{D}''$ is small enough, running $\text{iOFR}_{S}$ without any RCT methods may be faster. The figure below summarises Algorithm \ref{alg:NonSysId} in a flowchart. The following section will provide examples from the `NonSysId' package. 
\begin{figure}[h]
    \centering
    \includegraphics[scale=0.85]{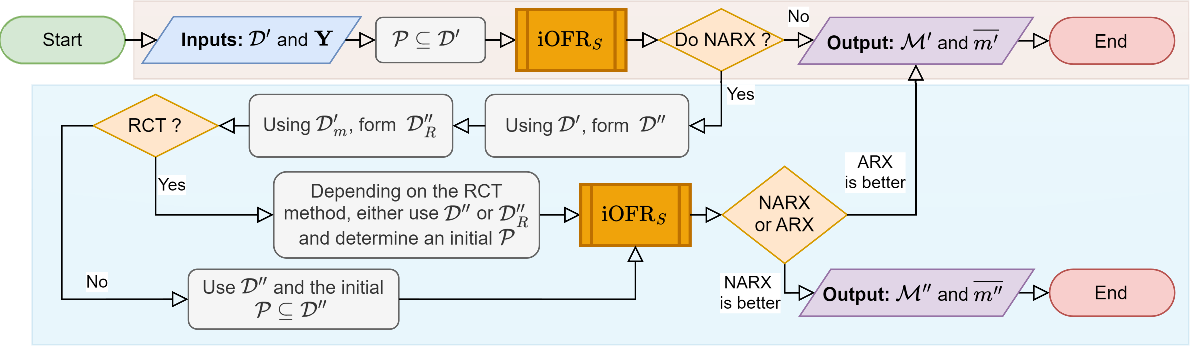}
    \caption{This flowchart summarises the procedures for identifying a (N)ARX model using $\text{iOFR}_{S}$ as described in Algorithm \ref{alg:NonSysId}. The region shaded in brown represents the ARX model identification process, while the blue-shaded region highlights the NARX procedures.}
    \label{fig:flowcharts}
\end{figure}

NARX models can be analysed in the frequency domain using Nonlinear Output Frequency Response Functions (NOFRFs) \cite{Lang2005}, which extend classical frequency response analysis to nonlinear systems \cite[Chapter~6]{billings2013a}. The NOFRF concept is an essential tool for system identification, describing how input frequencies interact nonlinearly to generate output frequencies that are harmonics and intermodulation effects. This facilitates a detailed understanding of how nonlinearities affect input-output dynamics \cite{Lang2005,BAYMA2018}. NOFRFs can be evaluated using various methods \cite{Gunawardena2018, ZHU2022}, providing enhanced insights into the frequency-domain behaviour of complex nonlinear systems. Consequently, NOFRFs enhance the utility of NARX models by offering a comprehensive framework for analysing and interpreting nonlinear system \cite{Zhu2021}. 

\section{Examples}
This section presents two examples showcasing the use of the NonSysID package, which implements $\text{iOFR}_{S}$ with PRESS-statistic-based term selection. The first example utilises synthetic data generated from a NARX model, while the second focuses on real data obtained from an electro-mechanical system.
%
\vspace{-1ex}
\subsection{Synthetic data example}
The following example demonstrates how to identify a NARX model using the `NonSysId' package. In this example, we consider a NARX model of a DC motor (Eq. \eqref{eq:NARX_eg}) as described in \cite{Lacerda2017}.
\begin{multline} \label{eq:NARX_eg}
    y(t) = 1.7813y(t-1) - 0.7962y(t-2) + 0.0339u(t-1) + 0.0338u(t-2)\\
    - 0.1597y(t-1)u(t-1) - 0.1396y(t-1)u(t-2)\\
    + 0.1297y(t-2)u(t-1) + 0.1086y(t-2)u(t-2) + 0.0085y(t-2)^2
\end{multline}
In Eq. \eqref{eq:NARX_eg}, $y(t)$ is the output and $u(t)$ is the input to the system at the time sample $t$. The NARX model is separately excited using two inputs: (a) White noise, where $u(t)\sim\mathcal{N}(0,1)$, and (b) a multi-tone sinusoidal wave defined as $u(t) = 0.2\big( 4\sin{(\pi t)} + 1.2\sin{(4\pi t)} + 1.5\sin{(8\pi t)} + 0.5\sin{(6\pi t)} \big)$. The model was simulated for 1000 time samples. Identification results for both input cases are presented below. Matlab scripts for this example are available in the code repository, along with documentation in the code repository provides a straightforward guide for using $\text{iOFR}_{S}$ in the `NonSysId' package.

\begin{figure}[!h]
    \centering
    \includegraphics[width=0.61\textwidth]{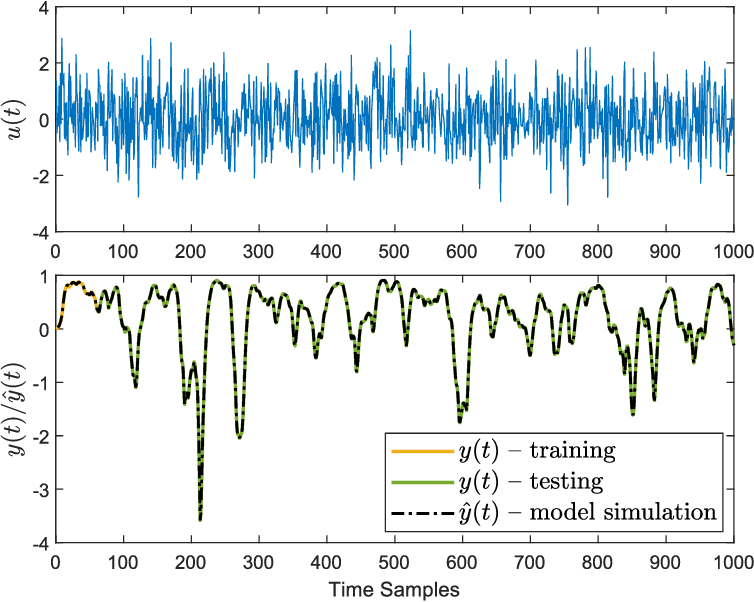}
    \caption{\textbf{Model identification results under input (a)}. The model simulation output $\hat{y}(t)$ is shown against the actual output $y(t)$ of the system given in Eq. \eqref{eq:NARX_eg}. The input $u(t)$ to the system is a Gaussian white noise signal $u(t)\sim\mathcal{N}(0,1)$. Only the first 60 samples are used for identifying/training the model using $\text{iOFR}_{S}$ in the `NonSysId' package. The variance of the error or model residuals in this case is $1.6018e^{-25}$}
    \label{fig:narx_eg_a_io}
\end{figure}

\begin{figure}[!h]
    \centering
    \includegraphics[width=0.61\textwidth]{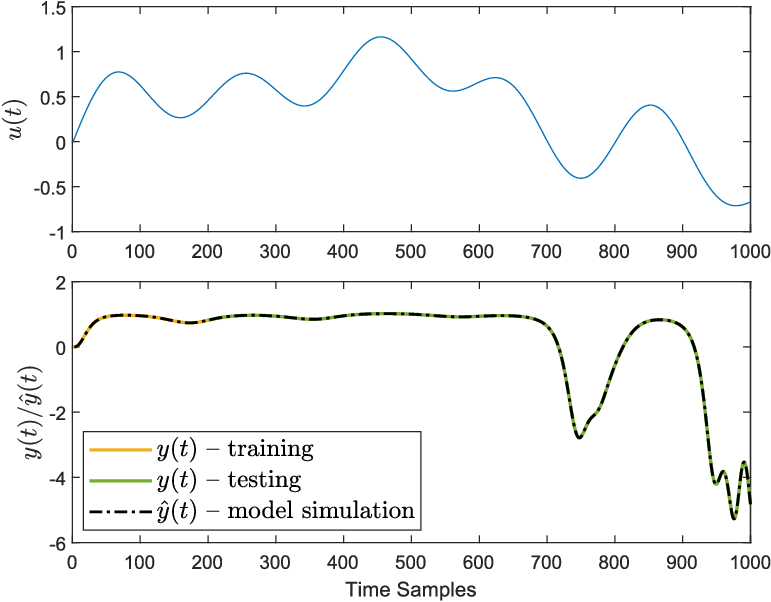}
    \caption{\textbf{Model identification results under input (b)}. The model simulated output, $\hat{y}(t)$, is compared with the actual output, $y(t)$, as defined in Eq. \eqref{eq:NARX_eg}. The input $u(t)$ to the system is a multi-tone sinusoidal signal given by $u(t) = 0.2\big( 4\sin{(\pi t)} + 1.2\sin{(4\pi t)} + 1.5\sin{(8\pi t)} + 0.5\sin{(6\pi t)} \big)$. In this case, the portion of $y(t)$ used for identification/training (yellow curve) is less informative compared to input (a), as fewer system dynamics are excited due to the limited frequency components in the input signal. Therefore, up to 200 samples are used for identifying the model using $\text{iOFR}_{S}$ in the `NonSysId' package. The variance of the error or model residuals in this scenario is $8.2178e^{-18}$. Using fewer than 200 samples results a sub-optimal model, as insufficient data limits the ability to capture the system's dynamics effectively.}
    \label{fig:narx_eg_b_io}
\end{figure}

Fig. \ref{fig:narx_eg_a_io} and \ref{fig:narx_eg_b_io} depict the training and testing data alongside the model simulated output for the inputs (a) and (b), respectively. The term `testing data' is used to refer to data not explicitly included during training, as the model is already validated through leave-one-out cross-validation during the identification/training process (see sub-section \ref{sec:PRESS}).

\begin{table}[!h]
    \normalsize
	\centering
	\caption{The model identified when Eq. \eqref{eq:NARX_eg} is excited with input (a), white noise}
	\vspace{1ex}
	\begin{tabular}{|c|c|c|c|}
		\hline
		Model term & \specialcell{Mean squared\\PRESS error} & ERR & Parameters/Coefficients\\
		\hline
        $y(t-1)$       & $1.342 \times 10^{-3}$   & $0.95001$                & \ \ $1.7813$ \\
        $y(t-2)$       & $1.6759 \times 10^{-4}$  & $2.255  \times 10^{-3}$  & $-0.7962$ \\
        $u(t-1)$       & $0.47871$                & $4.7434 \times 10^{-2}$  & \ \ $0.0339$ \\
        $u(t-2)$       & $6.8123 \times 10^{-5}$  & $1.8925 \times 10^{-4}$  & \ \ $0.0338$ \\
        $y(t-1)u(t-1)$ & $2.2653 \times 10^{-5}$  & $3.6489 \times 10^{-5}$  & $-0.1597$ \\
        $y(t-1)u(t-2)$ & $6.1439 \times 10^{-5}$  & $1.9004e \times 10^{-5}$ & $-0.1396$ \\
        $y(t-2)y(t-2)$ & $3.1515 \times 10^{-30}$ & $5.3837e \times 10^{-7}$ & \ \ $0.0085$ \\
        $y(t-2)u(t-1)$ & $3.7241 \times 10^{-7}$  & $2.9966e \times 10^{-5}$ & \ \ $0.1297$ \\
        $y(t-2)u(t-2)$ & $4.6109 \times 10^{-5}$  & $2.8901e \times 10^{-5}$ & \ \ $0.1086$ \\
    	\hline
	\end{tabular}%
	\label{tbl:inpt_a_param}%
\end{table}%

Tables \ref{tbl:inpt_a_param} and \ref{tbl:inpt_b_param} present the identified terms and parameter values of the corresponding NARX models under inputs (a) and (b), respectively. These tables also include the mean squared PRESS error and the ERR metrics for each term. The values of these metrics depend on the order in which the terms were added to the model during the forward selection procedure, determined by the orthogonalization path taken by the OFR algorithm (sub-section \ref{sec:iOFR}). The mean squared PRESS error reflects the one-step-ahead leave-one-out cross-validation error after the term is added to the model. Sorting tables \ref{tbl:inpt_a_param} and \ref{tbl:inpt_b_param} in descending order of the mean squared PRESS error reveals the sequence of the terms added. For example, in table \ref{tbl:inpt_a_param}, the term $u(t-1)$ was added first (indicating the orthogonalization path starts with this term) followed by $y(t-1)$, $y(t-2)$, and so on. The ERR represents the proportion of the actual output variance (variance of $y(t)$) explained by each corresponding term.
\begin{table}[!h]
    \normalsize
	\centering
	\caption{The model identified when Eq. \eqref{eq:NARX_eg} is excited with input (b), a multi-tone sinusoid}
	\vspace{1ex}
	\begin{tabular}{|c|c|c|c|}
		\hline
		Model term & \specialcell{Mean squared\\PRESS error} & ERR & Parameters/Coefficients\\
		\hline
        $y(t-1)$       & $1.2209 \times 10^{-4}$    & $0.1035$                  & \ \ $1.7813$  \\
        $y(t-2)$       & $7.0858 \times 10^{-7}$    & $1.7841  \times 10^{-4}$  & $-0.7962$ \\
        $u(t-1)$       & $2.8085 \times 10^{-9}$    & $2.5768 \times 10^{-9}$   & \ \ $0.0339$  \\
        $u(t-2)$       & $3.7183 \times 10^{-8}$    & $3.5856 \times 10^{-7}$   & \ \ $0.0338$  \\
        $y(t-1)u(t-1)$ & $4.5778 \times 10^{-12}$   & $2.7792 \times 10^{-9}$   & $-0.1597$ \\
        $y(t-1)u(t-2)$ & $2.9234 \times 10^{-7}$    & $6.0493 \times 10^{-7}$   & $-0.1396$ \\
        $y(t-2)y(t-2)$ & $3.8123 \times 10^{-9}$    & $4.6086 \times 10^{-8}$   & \ \ $0.0085$  \\
        $y(t-2)u(t-1)$ & $1.9182 \times 10^{-25}$   & $6.4198 \times 10^{-12}$  & \ \ $0.1297$  \\
        $y(t-2)u(t-2)$ & $7.0559 \times 10^{-2}$    & $0.89632$                 & \ \ $0.1086$  \\
    	\hline
	\end{tabular}%
	\label{tbl:inpt_b_param}%
\end{table}%
\begin{figure}[!htb]
    \centering
    \includegraphics[width=0.66\textwidth]{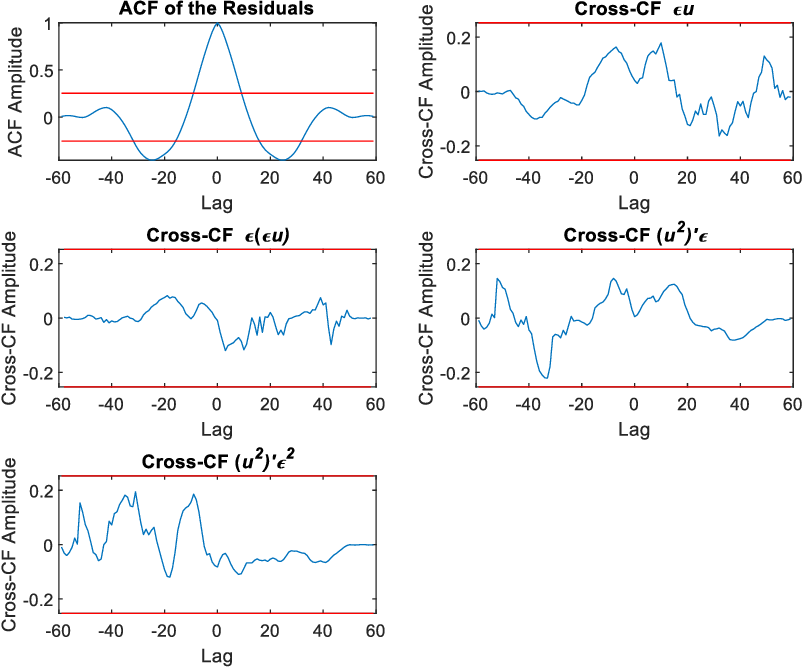}
    \caption{Model validation results for input (a). The red bounds indicate the tolerances the correlation function should stay within for the identified model to be unbiased.}
    \label{fig:narx_eg_a_val}
\end{figure}
\begin{figure}[!htb]
    \centering
    \includegraphics[width=0.66\textwidth]{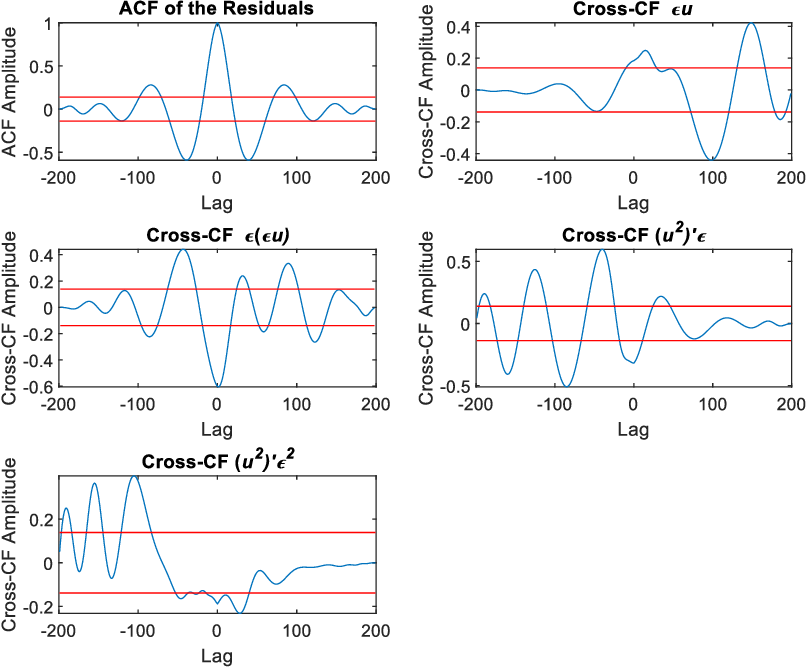}
    \caption{Model validation results for input (b). The red bounds indicate the tolerances the correlation function should stay within for the identified model to be unbiased.}
    \label{fig:narx_eg_b_val}
\end{figure}

The correlation based statistical validation tests for nonlinear models \cite{Billings1983} are presented in Fig. \ref{fig:narx_eg_a_val} and \ref{fig:narx_eg_b_val}. These validation tests are conducted on the training data (yellow region of $y(t)$ in Fig. \ref{fig:narx_eg_a_io} and \ref{fig:narx_eg_b_io}). From the auto-correlation function (ACF) of the residuals, it is observed that the model residuals, in both cases (a) and (b), are not entirely white noise. Additionally, in Fig. \ref{fig:narx_eg_b_val}, the cross-correlation functions (Cross-CF) between the input $u(t)$ and the model residuals are not completely within the tolerance bounds, indicating some bias in the model. However, the variance of the model residuals are $1.6018e^{-25}$ and $8.2178e^{-18}$, respectively, for (a) and (b), compared to the training data variances of $0.069$ and $0.0581$. This shows that the bias of the identified model is minimal. As such, even though the identified terms and parameters (Tables \ref{tbl:inpt_a_param} and \ref{tbl:inpt_b_param}) are similar to the actual system (Eq. \eqref{eq:NARX_eg}), the parameters do have differences considering from the 4\textsuperscript{th} decimal place and beyond.
\FloatBarrier
\subsection{Real data example}
The real data in this example is obtained from an electromechanical system described in \cite{Lacerda2017b}. The system comprises two 6V DC motors mechanically coupled by a shaft. One motor acts as the driver, transferring mechanical energy, while the other operates as a generator, converting the mechanical energy into electrical energy. The system input is the voltage applied to the DC motor acting as the driver. This input is a pseudo-random binary signal (PRBS) designed to excite the system over a range of dynamics. The output of the system is the rotational speed (angular velocity) of the generator motor.

\begin{figure}[!htb]
    \centering
    \includegraphics[width=\textwidth]{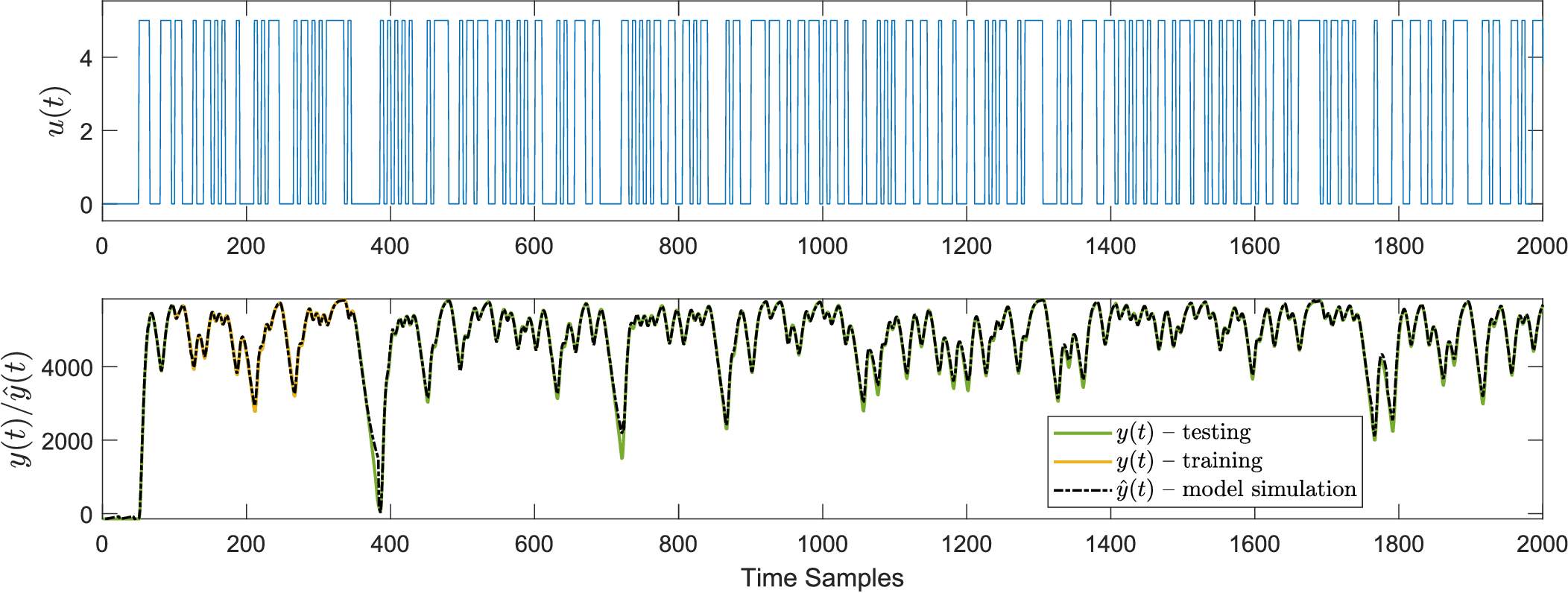}
    \caption{\textbf{Model identification results from the electro-mechanical system}. The model simulation output $\hat{y}(t)$ is presented against the actual output $y(t)$ of the system given in Eq. \eqref{eq:NARX_eg}. The input $u(t)$ to the system is a PRBS. Only 250 samples are used for identifying/training the model using $\text{iOFR}_{S}$ in the `NonSysId' package.}
    \label{fig:narx_eg_rldt_sys}
\end{figure}
\begin{figure}[!htb]
    \centering
    \includegraphics[width=0.8\textwidth]{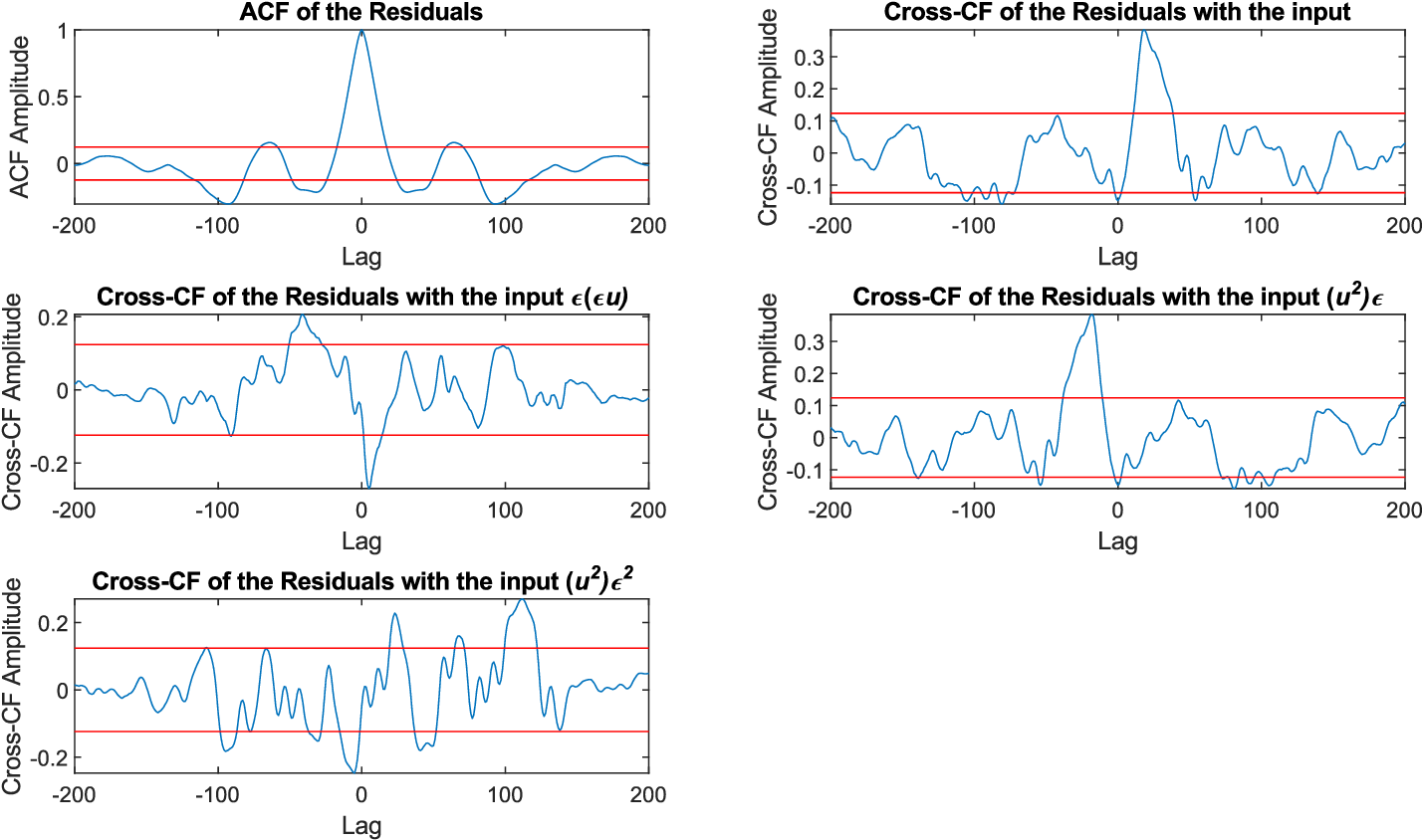}
    \caption{Model validation results for the system in \cite{Lacerda2017b}. The red bounds indicate the tolerances the correlation function should stay within for the identified model to be unbiased.}
    \label{fig:narx_eg_rldt_val}
\end{figure}
\begin{table}[!h]
    \normalsize
	\centering
	\caption{The model identified from the data generated from the system in \cite{Lacerda2017b}}
	\vspace{1ex}
	\begin{tabular}{|c|c|c|c|}
		\hline
		Model term & \specialcell{Mean squared\\PRESS error} & ERR & Parameters/Coefficients\\
		\hline
        $y(t-1)$       & $8128.5$              & $0.49526$              & $1.7844$   \\
        $y(t-2)$       & $975.85$              & $0.00028497$           & $-0.79156$ \\
        $u(t-1)$       & $318.88$              & $2.6363\times 10^{-5}$ & $47.205$   \\
        $y(t-2)u(t-1)$ & $158.23$              & $6.211\times 10^{-6}$  & $-0.037612$\\
        $y(t-3)u(t-1)$ & $1.2306\times 10^{7}$ & $0.50441$              & $0.030086$ \\
        $u(t-2)u(t-2)$ & $91.271$              & $2.5147\times 10^{-6}$ & $1.89$     \\
        $u(t-2)u(t-3)$ & $71.842$              & $7.2261\times 10^{-7}$ & $-0.91694$ \\
    	\hline
	\end{tabular}%
	\label{tbl:narx_eg_rldt_val}%
\end{table}%

\FloatBarrier
\section{Future Work}
Currently, the `NonSysId' package is capable of identifying single-input single-output (SISO) and multi-input single-output (MISO) models. However, the correlation-based residual analysis is limited to handling only SISO models. In the future, we plan to extend the package to identify multi-input multi-output (MIMO) and enable validation for both MISO and MIMO systems. In \cite{CHEN2006}, a local regularisation method for the OFR was introduced. This will be incorporated into $\text{iOFR}_{S}$. While the `NonSysId' package currently supports polynomial NARX models, future versions will broaden its scope to allow  $\text{iOFR}_{S}$ to be applied to a wider range of basis functions. Furthermore, an open-sourced Python version of this package is expected to be released in the future.

\bibliographystyle{ieeetr}
\bibliography{citations.bib}

\end{document}